\documentstyle[12pt]{article}
\newcommand{\bold}[1]{\mbox{\boldmath $#1$}}    
\newcommand{\ie}{{\em i.e.}}                    
\newcommand{\etal}{{\em et al.}}                
\newcommand{\MeV}{{\rm MeV}}                    
\newcommand{\fm}{{\rm fm}}                      
\newcommand{\r}{{\bold{r}}}                     
\newcommand{\del}{\partial}                     
\newcommand{\pphi}{{\bold{\phi}}}		
\newcommand{\beq}{\begin{equation}}
\newcommand{\eeq}{\end{equation}}
\newcommand{\bfig}{\begin{figure}}
\newcommand{\efig}{\end{figure}}
\newcommand{\DWT}{{\sl DWT}}			
\newcommand{\DCC}{{\sl DCC}}			

\begin{document}
\begin{titlepage}

\noindent{\sl Physical Review D}\hfill LBNL-40184; AZPH-TH/97-06\\[8ex]

\begin{center}
{\large {\bf Analysis of DCC Domain Structure$^*$}}\\[8ex]
{\sl J\o rgen  Randrup}\\[1ex]
Nuclear Science Division, Lawrence Berkeley National Laboratory\\
University of California, Berkeley, California 94720\\[2ex]
and\\[2ex]
{\sl Robert L. Thews}\\[1ex]
Department of Physics, University of Arizona,
Tucson, Arizona 85721
\\[6ex]
May 7, 1997\\[6ex]
{\sl Abstract:}\\
\end{center}

{\small\noindent
Wavelet-type methods are employed for the analysis of pion field configurations
that have been obtained by dynamical simulations in idealized scenarios
relevant to the formation of disoriented chiral condensates.
It is illustrated how the measurement of the isospin domain structure
depends on the ability to zoom in on limited parts of the phase space,
due to the interplay between the pion correlation length
and the effective source geometry.
The need for advanced analysis methods is underscored by the fact that
the extracted neutral-fraction distribution
would differ significantly from the ideal form,
even under perfect experimental conditions,
and, moreover, by the circumstance that
thermal sources with suitably adjusted temperatures
can lead to distributions that may be practically indistinguishable
from those arising from \DCC-type non-equilibrium evolutions.\\[6ex]


\vfill
\noindent
$^*$This work was supported by the Director, Office of Energy Research,
Office of High Energy and Nuclear Physics,
Nuclear Physics Division of the U.S. Department of Energy
under Contract No.\ DE-AC03-76SF00098.\\
\noindent
}
\end{titlepage}

\section{Introduction}
\label{introduction}

The prospect of producing disoriented chiral condensates
in high-energy collisions of hadrons or nuclei
has stimulated significant interest over the past few years.
The \DCC\ phenomenon is expected to occur
as a result of the non-equilibrium relaxation of the chiral field
towards its normal vacuum values, starting from a phase
in which chiral symmetry is temporarily restored.
\cite{Anselm88,Anselm91,Blaizot:PRD46,Rajagopal:NPB399,Bjorken}.
This evolution may amplify soft pion modes,
so that the isospin field becomes correlated over large distances,
leading to observable anomalies in the pion multiplicity distributions.
However, it is inherently difficult to observe the phenomenon
and it is therefore important to develop suitable methods of analysis.

The present paper reports on the application of wavelet-like analyses
to field configurations that have been obtained by dynamical calculations
with the linear sigma model.
The possibility of applying wavelet techniques for the analysis
of \DCC\ domain structure was recently suggested by Huang \etal \cite{HSTW:95}
and instructive schematic applications were made.
The present work extends this pioneering effort in two ways:
1) we further develop the analysis methods so that more realistic situations
can be addressed, such as three-dimensional pion field configurations, and
2) we apply the various analyses to more realistic scenarios containing
pion field configurations that have been generated by dynamical simulations
of the conditions expected to occur in the course of actual collisions.

\section{Scenarios}

The earlier wavelet analyses \cite{HSTW:95}
have been performed for one-dimensional systems.
In the present study we wish to address scenarios
in which the underlying pion fields are fully three-dimensional
and have been generated by non-equlibrium evolutions
of the kind characteristic of the \DCC\ phenomenon.

For this purpose,
we employ the linear $\sigma$ model at the semi-classical level,
as described in Ref.\ \cite{JR:PRD}.
The chiral degrees of freedom are then described by a real $O(4)$ field,
$\pphi(\r,t)=(\sigma,\bold{\pi})$,
and subject to a quartic self-interaction,
$V={\lambda\over4}(\phi^2-v^2)^2-H\sigma$.
The field is confined within a rectangular box,
with periodic boundary conditions imposed,
and is generally sampled from a thermal equilibrium ensemble,
using the quantal Bose-Einstein occupation numbers.
In this manner it is ensured that the field
has a realistic correlation function.
In the present study,
the field is represented on a cartesian grid in position space
with a spacing of ${1\over8}\ \fm$.
We consider two different overall geometries
which we shall now describe in turn.

In order to very roughly mimic scenarios that may be encountered
in high-energy nuclear collisions,
we consider first ``rope'' configurations
in which the confining box is 32 fm long and has a cross section of
$\Sigma_0=4$$\times$$4\ \fm^2$.
These configurations can be thought of as representing the result of the
string decays in the central rapidity region
in the wake of a high-energy collision,
with the $z$ coordinate corresponding to the rapidity.
The rope system can thus be regarded as 256 (=$32$$\times$$8$)
square-shaped slices of matter,
each one having a thickness of ${1\over8}\ \fm$.
The basic data set which we wish to subject to analysis
is obtained by summing the pion yields over the transverse directions,
\beq\label{sum}
n_i(z)\ \equiv\ \int {dx dy \over \Sigma_0} \pi_i(x,y,z)^2
\eeq
for each of the three cartesian directions of the pion field,
$i=1,2,3$.
Due to the discretization,
these quantities can be enumerated by the index of a given slice, $\kappa$,
and the integral over the cross section of the rope
is a double sum over grid points.
For the sake of the analysis,
we shall assume that the yield of neutral pions from a given slice $\kappa$ 
is proportional to the associated value of $n_3^{(\kappa)}$ and
that the total pion yield from that slice is
$n^{(\kappa)}=n_1^{(\kappa)}+n_2^{(\kappa)}+n_3^{(\kappa)}$.
In a more refined description,
the time derivative of the field should be taken into account as well
when extracting the pion contents but this complication
has little practical import on the present study,
due to the transverse averaging over the 1024 individual grid points
for each slice.
The important thing is to generate a data sample with some resemblence
to what could be expected in realistic scenarios.

As an alternative geometry,
we consider the confining box to be an $8$$\times$$8$$\times$$8\ \fm^3$ cube,
which may be taken to mimic a large fireball.
These ``cube'' configurations are relevant for central collisions
of large nuclei where the volume within which
chiral symmetry is temporarily restored is considerable.
While the the total volume is the same for the cube and rope configurations,
the basic data set for the cube geometry is obtained by
averaging the pion yields over a $1$$\times$$1$$\times1$$\ \fm^3$ cubic cell
(which contains a total of $8^3$=512 grid points),
in analogy with (\ref{sum}).

For both types of geometry,
we consider samples of field configurations
obtained for various dynamical scenarios.
Primarily,
we wish to consider correlated fields of the type
that would be expected as a result of the non-equilibrium evolution
following the early approximate restoration of chiral symmetry,
in which the soft pion modes experience significant amplification.
In order to generate semi-realistic samples,
we employ the method of pseudo-expansion introduced in Ref.\ \cite{JR:PRL}.
The initial field configuration is then sampled from
a thermal equilibrium ensemble at a suitable high temperature
(as in Ref.\ \cite{JR:PRL}, we use $T_0=400\ \MeV$)
and then subjected to a rapid cooling
by means of a Rayeligh dissipation term.
The corresponding equation of motion is then
\beq\label{EoM}
[\Box +\lambda(\phi^2-v^2)]\pphi-H\hat{\sigma}=-{D\over t}\del_t\pphi\ .
\eeq
Roughly speaking,
this equation mimics the effect of a Bjorken-type scaling expansion
in $D$ dimensions.
It was shown in Ref.\ \cite{JR:PRL} that $D$=1
(corresponding to the familiar longitudinal scaling expansion
\cite{Huang:PRD49,Kluger})
is too slow to generate any amplification,
whereas $D$=3 (corresponding to an isotropic expansion \cite{GM,Lampert})
leads to substantial growth of the soft pion modes.
We shall consider fields that have been obtained with
intermediate ($D$=2) and fast ($D$=3) cooling,
in which cases significant non-equilibrium behavior occurs \cite{JR:PRL}.

In addition to these non-equilibrium scenarios,
we shall also consider samples of rope and cube configurations
drawn from thermal equilibrium ensembles \cite{JR:PRD}.

\section{Wavelet formalism}

{}From the spatial dependence of the pion charge fractions in the 
events generated, we wish to extract information about possible
domains of \DCC.  These domains will obviously occur with various
scales and be located at various positions in space.  Hence we would
like to perform an analysis which simultaneously isolates candidate
structure in both space and scale.  A technique well-suited to this
task has been developed over the past several decades, known as
the discrete wavelet transform (\DWT), or more generically as wavelet
analysis. Applications of this technique have previously proved
useful in detecting structures on various scales in turbulence 
\cite{FARGE:92}, multiparticle production \cite{GREINER:95,GREINER:96},
and astrophysics \cite{FANG:95,FANG:96}.  We refer the reader to
several standard surveys and introductory treatments of the \DWT\ for
technical details \cite{DAUB:92,CHUI:92,MEYER:93,KAISER:94}.  

The essential feature of the wavelet formalism is that one can find
complete orthonormal  sets of wavelet (father) functions $\psi_{jk}(z)$
and scaling (mother) functions $\phi_{jk}(z)$ which are localized
both in space and in scale.
The index $j$ labels the scale of the basis function and ranges
in integer steps from $0$ to $j_{\rm max}$, such that the finest
resolution for some signal function 
$f(z)$ corresponds to a histogram with $2^{j_{\rm max}}$ bins.
The index $k$ labels the spatial bin for a given $j$
and ranges from $0$ to $2^{j}$-$1$.
Hence each step in $j$ leads to a change by a factor of $2$ in scale.

The wavelet functions $\psi_{jk}(z)$ are orthogonal with respect to
all $j,k$, such that in the expansion for a given scale $j$, 
\begin{equation}\label{father}
\tilde{f}^{(j)}(z)=\sum_{k=0}^{2^{j}-1} \tilde{f}_{jk}\psi_{jk}(z)\ ,
\end{equation}
the expansion coefficients can be obtained directly by projection,
\begin{equation}
\tilde{f}_{jk}=2^j\int f(z)\psi_{jk}(z)dz\ .
\end{equation}
Thus the father-function coefficients $\tilde{f}_{jk}$
provide a view of the function $f(z)$ at the single scale $j$.  

By contrast,
the scaling functions $\phi_{jk}(z)$ are orthogonal only within a
given $j$ subspace.
Considering the corresponding scaling-function expansion,
\begin{equation}\label{mother}
f^{(j)}(z)=\sum_{k=0}^{2^{j}-1}f_{jk}\phi_{jk}(z)\ ,
\end{equation}
one can see that the mother-function coefficients $f_{jk}$
contain structure present both at the given scale $j$
and at all coarser scales (lower $j$ values).
Indeed, it is readily shown that the two representations are related as
\begin{equation}
\tilde{f}^{(j)}(z)=f^{(j+1)}(z)-f^{(j)}(z)\ .
\label{EQW}
\end{equation}
The interpretation is quite simple: structure at a give scale is
characterized by fluctuations in going to the next finer scale in resolution,
such that the entire multiresolution expression for $f(z)$
is simply a sum of all wavelet function expansions.
The corresponding scaling (mother) function expansion 
contains all fluctuations for scales from $j$=$0$ up to some given $j$.
The mother-function coefficients $f_{jk}$ can be obtained by recursion relations
relating them to the father function coefficients.
It is even simpler to produce a mother function representation of $f(z)$
at a given scale by truncating the wavelet function expansion (\ref{father})
at the corresponding $j$.

It is not a trivial matter in general to construct wavelet and scaling
functions with the desired locality in both space and scale.  However, one
such set is quite well-known: the Haar wavelets.
The associated scaling functions have the form of an elementary square wave,
with $z$ scaled to the unit interval [0,1],
\begin{eqnarray}
\phi_{jk}(z)=\left\{ \begin{array}{ll}
                        1 & \mbox{$k\ 2^{-j}\leq z < (k+1)\ 2^{-j}$} \\
                        0 & \mbox{otherwise}
                        \end{array} \right.\ ,
\end{eqnarray}
while the wavelet functions are elementary bipolar square waves,
\begin{eqnarray}
\psi_{jk}(z)=\left\{ \begin{array}{ll}
	\phantom{-}1 &
	\phantom{mmm}\mbox{$k\ 2^{-j}\leq z < (k+{1\over2})\ 2^{-j}$} \\
                 -1 & \mbox{$(k+{1\over2})\ 2^{-j}\leq z < (k+1)\ 2^{-j}$} \\
       \phantom{-}0 & \mbox{otherwise}
                        \end{array} \right. .
\end{eqnarray}
The interpretation of these expansions is then obvious:
The mother-function expansion (\ref{mother}) is just a binned histogram of
the function with bin width given by the scale $j$ and position given by $k$,
while the father-function expansion (\ref{father}) contains information
on the differences between adjacent bins (in $k$) for each scale $j$.

These basis functions satisfy all requirements
but are unfortunately discontinuous.
Thus the scale information is not well localized
but must recieve support in an infinite domain in
order to build up the discontinuity.
For most of our analysis,
this is not a serious defect.
However,
we would also like to look at the scale dependence of the power spectrum.
For this purpose,
we will also use a set of wavelet functions first employed by
Daubechies \cite{DAUB:92}.
These functions are both localized and continuous,
and are denoted by {\sl DAUBn}, with $n = 4,6,\dots,20$.
Details of their construction are given in Ref.\ \cite{PRESS:92}.

Analagous to Fourier analysis, a
Parseval Theorem for the \DWT\ utilizes the complete and orthonormal
properties of the father function basis functions $\psi_{jk}(z)$:
\begin{equation}
\int_0^1|f(z)|^2dx=\sum_{j=0}^{\infty}\frac{1}{2^j}
\sum_{k=0}^{2^j-1}|\tilde{f}_{jk}|^2\;\ .
\end{equation}
Thus, one can define the power spectrum
with respect to the wavelet basis at each scale $j$,
\begin{equation}
W_j=\frac{1}{2^j}\sum_{k=0}^{2^j-1}|\tilde{f}_{jk}|^2\; .
\end{equation}
The quantity $W_j$ is then the power of fluctuations
on a scale which is a fraction $2^{-j-1}$ of the overall length
(since $\tilde{f}_{jk}$ first contributes to structure in $f^{(j+1)}$)
and it contains information {\it only} about this single scale.
Results of such an analysis are presented in the next section. 

\section{Applications}

For each scenario considered,
we generate a sample of 100 field configurations,
in order to achieve a reasonable statistical significance.
For each configuration in the sample,
we first calculate the neutral pion fraction, $f=n_3/n$,
and generate a histogram of the corresponding distribution, $P_9(f)$.
In order to analyze the scale dependence of the observables
extracted from the rope configurations,
we generate a hierarchy of data sets by successive subdivision.
Thus, we divide each rope configuration into two equal parts
and repeat the extraction procedure for the resulting 200 sources,
yielding a histogram of $P_8(f)$.
This process of equal bisection is continued
until we reach the basic level of a single slice,
at which level we have a total of $100$$\times$$256$ individual sources,
yielding $P_1(f)$.
The notation $P_L(f)$ is chosen to facilitate comparison of distributions
from both rope and cube geometries with the same total volume, 
$\Omega = 2^L\ \fm^3$.

It is clear that the procedure for obtaining the distributions $P_L(f)$
is equivalent to performing a mother-function expansion at various scales
using the Haar wavelet, with $j = 9-L$.
Such a procedure and its characterization in terms
of the distribution histograms were first presented in Ref.\ \cite{HSTW:95}.
(We note that here the equivalent $L$ value has a maximum of $8$ in the 
wavelet formalism, since one does not generally utilize the average
value of the signal function as a separate step in the wavelet
hierarchy, \ie\ the data sets for $j$=$j_{\rm min}=1$ contain $2^j = 2$ bins).
The analysis was also performed using the {\sl DAUB4} and {\sl DAUB20} wavelets,
and although the event shapes were not identical in detail, there were no
signicant differences in the distributions $P_L(f)$.

In order to characterize the scaling behavior in a fairly compact manner,
we extract the width of the neutral pion fraction distribution, $\Delta f$,
\beq
(\Delta f)^2\ =\ <f^2> - <f>^2\ = \int df\ f^2\ P(f) - (\int df\ f\ P(f))^2\ .
\eeq
We note that the fraction is one third on the average,
$<f>={1\over3}$.
As a reference, we consider what would result
if each basic slice were a fully coherent source
with a randomly oriented field vector of constant magnitude.
Then $P_1(f)$ would be equal to the ideal form $1/2\sqrt{f}$
and the distributions for the coarser scales would be the
corresponding multiple convolutions (see Ref.\ \cite{JR:NPA}).
In particular, the width would be given by
$\Delta f =(2^{3-L}/45)^{1/2}$.

In order to analyze the scale dependence of the cube configurations,
we proceed in analogy with the rope treatment
and first extract the neutral pion distribution corresponding to the
coarsest scale, $P_9(f)$,
by treating each of the 100 configurations as a single source of pions.
Subsequently,
we subdivide the cube into eight subcubes.
This changes the scale by three levels and thus yields $P_6(f)$.
A further ocotopartition yields $P_3(f)$,
and a final partition into $100$$\times$$512$ sources
of volume $1$$\times$$1$$\times$$1\ \fm^3$ yields  $P_0(f)$.
Again, this procedure is equivalent to a 3-dimensional mother function
representation in the Haar basis,
restricted to equal scales $j=(9-L)/3$ in each dimension.

\section{Results}

The rope events are essentially one-dimensional,
though they have been obtained with a fully three-dimensional calculation,
and so they can be readily subjected to the same wavelet analyses
as were made in Ref.\ \cite{HSTW:95}.
However,
whereas the data sets employed there were generated from fields
whose values were totally uncorrelated from one grid point to the next,
the present data sets have been generated with a realistic correlation length
built in from the beginning and, consequently,
they would be expected to be somewhat more realistic
(although they are still very idealized, of course).

As an illustration of this feature,
we show in Fig.\ \ref{fig:f} a histogram of the neutral pion fraction $f$
for two rope events that were obtained with $D$=3.
It is clearly seen how the value of $f$ changes in a relatively smooth fashion
as one moves from one slice to the next,
in reflection of the finite correlation length in the underlying pion fields.
These two events were chosen to illustrate the typical range
of scale size for the ``domains'' and it should be noted that
the domain structure exhibits relatively large fluctuations from event to event,
suggesting that a systematic statistical analysis is needed.

Figure \ref{fig:Pf-rope} shows the result of subjecting the rope ensembles
to the bisection (or Haar wavelet) analysis described above.
In order not the clutter the display,
only results for some of the scales have been shown.
The extracted distributions of the neutral fraction, $P_L(f)$,
exhibit a gradual change as the scale of analysis is changed
from very coarse to very fine.
The coarsest scale,
when the entire event is treated as a single source,
is much larger than the correlation length.
The corresponding distribution is well peaked around one third,
the limiting value obtained for large source volumes.
As the repeated bisection reduces the size of the individual sources,
the isospin vectors of the pions associated with a given source
become increasingly aligned and, consequently,
the neutral-fraction distribution $P_L(f)$ steadily broadens.
The anomaly,
\ie\ the deviation from a well-peaked distribution around $1\over3$,
is generally larger for the sample that has has been most rapidly cooled,
$D$=3, as is to be expected
because those events have experienced the most pronounced growth of
the soft modes and thus possess the longest correlation length \cite{JR:NPA}.
The difference between the two cooling rates
is evident already at the coarsest scale
and it grows steadily more pronounced.
However, for each of the two cooling scenarios
the distribution $P_L(f)$ approaches a limiting form
as the scale is reduced and, furthermore,
these limiting forms still differ significantly
from the idealized $1/2\sqrt{f}$ form.

This important feature is a consequence of the fact
that the ``observed'' pion yields are obtained
by summing over the cross section of the rope,
for each individual slice,
corresponding to what would be done experimentally
if the measured pions were binned according to rapidity only.
This averaging procedure leads to a degradation of the signal,
since the pion field may change its direction significantly
over the $4$$\times$$4$~fm$^2$ transverse cross section of a given slice,
even if the field changes only slightly 
as one moves through its $1\over8$-fm thickness in the longitudinal direction. 
Indeed,
when the correlation length is smaller that the width of the rope,
several independent sources effectively contribute and
the resulting distribution is a result of a multiple convolution
of the idealized distribution corresponding to a fully isospin aligned source
(see Ref.\ \cite{JR:NPA}).
Since the correlation length is larger for the $D$=3 events,
the pion field maintains its direction better across any given slice and
the effective number of sources per slice is then smaller for for $D$=2.
This fact is manifested in a significantly more anomalous form of $P(f)$.
In fact, making a comparison with Fig.\ 3 in Ref.\ \cite{JR:NPA},
it appears that the $P(f)$ extracted for $D$=3 corresponds roughly
to that obtained analytically for two independent sources
in the scenarios with a random overall multiplicity,
whereas the distribution extracted for $D$=2 looks more like the
one obtained for six sources.
These observations are quite compatible with the fact that the
correlation lengths,
as measured by the full width at half maximum of the pion correlation function,
are approximately 2.8 fm and 1.6 fm, for $D$=3 and $D$=2, respectively
(as extracted from Fig.\ 2 in Ref.\ \cite{JR:NPA}).
One would then expect $(2.8/1.6)^2\approx3$ more sources per area for $D$=2,
which indeed appears to be the case, as just noted (six versus two).
Moreover, for $D$=3 one might expect that a square with a side length of 4 fm
would be able to accommodate roughly $(4/2.8)^2\approx2$ sources
(and analogously $(4/1.6)^2\approx6$ sources for $D$=2)
which is also in good accordance with what we just noted.

In Fig.\ \ref{waveletwidth} we display the scale-dependence of the
width of these distributions, $\Delta f$.
The individual events were analyzed
using mother function {\sl DAUB4} wavelet decompositions.
A reference distribution was produced by using a randomly oriented
pion field vector in each elementary slice of the rope events.
The corresponding widths of these reference 
distributions at each scale
are shown by the dotted curve,
which coincides with the expectation based on multiple convolutions
of the ideal $1/2 \sqrt{f}$ distribution starting at the smallest scale.  
For the $D$=2 and $D$=3 events,
an increase of $\Delta f$ is also seen with reduction of scale size,
but the rise saturates
as expected at scales much smaller than the correlation length.
The region of most rapid growth is seen for $4 \leq L \leq 7$,
corresponding to physical scales between 1-8 fm,
with a more rapid change for the $D$=3 fast expansion scenario.

An important lesson that may be learned from the above analysis
of the rope events is that any phase-space averaging
is likely to degrade the signal.
Specifically,
the fact that all pions near a given value of $z$ are lumped together
tends to eliminate any coherence that might have been present in the
underlying field.
Experimentally,
this would translate into an expectation that binning the pions only
with respect to rapidity might not be sufficiently revealing.

In order to explore this potentially important aspect further,
we now turn to the alternate geometry where the source is cubic in shape
and the subdivisions are made in a manner that does not discriminate
one direction from another.
Figure \ref{fig:Pf-cube} shows the resulting distributions
of the neutral pion fraction $P(f)$ obtained at the various scales
and should be compared with the results for the corresponding rope samples,
Fig.\ \ref{fig:Pf-rope}.
We note that as long as the source sizes are suitably large,
the two different geometries yield rather similar results,
with the main determining factor being the overall volume of the source
at the scale considered, $\Delta\Omega$.
However, important differences begin to appear
once the spatial extension of the indivudual source
becomes comparable to the pion correlation length.
In particular,
the distribution $P(f)$ will now approach the ideal form,
because the cubic sources are being shrunk in all directions uniformly
and so the fields within each will become increasingly aligned
and there will be no degradation due to contributions from misaligned
isospin orientations,
contrary to the case of the slices discussed above.
Again the effect is strongest for $D$=3,
where the distribution extracted for the smallest cube (having volume 1 fm$^3$)
is now rather close to $1/2\sqrt{f}$.

To facilitate the comparison of these scenarios,
we show in Fig.\ \ref{fig:DfV} the width $\Delta f$ of the distributions
$P(f)$ extracted for the various samples,
as a function of the source size used.
This display shows clearly how the results extracted from
the rope samples saturate,
whereas those obtained for the cubes keep growing steadily.
If the subdivision had been continued below the scale of 1 fm$^3$,
the width would ultimately saturate at the value for idealized distribution,
$\Delta_0=2/\sqrt{45}\approx0.30$.
However,
such small sources have not been considered,
because a source of volume 1 fm$^3$ contains only a few pions
and much smaller sources would thus be physically meaningless.
It is thus not realistic to expect that an actual observation
could yield the idealized distribution,
even under perfect experimental conditions,
and one can at best expect distorted distributions of the forms
shown in Figs.\ \ref{fig:Pf-rope} and \ref{fig:Pf-cube}. 
Moreover,
our analysis clearly suggests that
in order to optimize the appearance of the signal
(\ie\ the occurrence of an anomalous distribution),
it is important to avoid combining pions from different domains.
Thus a careful data binning is required,
so as to zoom in on a single domain.
For example,
one might imagine a procedure in which a discrimination of the observed pions
is made not only on the basis of their rapidity
but also with regard to azimuthal angle and energy.
Even then, the pions contributing to a single bin
might arise from different locations in the system
and thus, presumably, be uncorrelated.
Naturally,
the lower the energy that can be measured,
the larger the corresponding correlation length,
and thus the larger the rapidity and azimuthal bin sizes
that can be tolerated.

The scale dependence of the structure may be examined further by employing
the father-function representation to extract the wavelet power spectrum.
This representation has the virtue of completely isolating structure
at single scale $j$ of the wavelet basis.
We calculate the power spectrum $W_j$ for each rope configuration,
and then average over the 100 events
to get an average and standard deviation at each $j$.
The results are shown in Fig.\ \ref{waveletpower}.
The dotted line is obtained from input data
generated by assigning random numbers to the pion charge yields
at each basic grid point.  
Its small magnitude is indicative of the averaging
process over the cross section of the rope geometry, which suppresses any
large fluctuations of the signal at the smallest scales,
as discussed above.
The flatness of the curve verifies that any structure is truly random in nature,
\ie\ it has no dependence on scale.
The corresponding power spectra for the $D$=2 and $D$=3 events
start out at a very low level for small scales,
but then exhibit a very strong growth at larger scales.
They reach an approximate plateau at $1 \leq j \leq 4$,
which corresponds to the same scales of $1-8$ fm
that were found to be of significance in the mother function analysis.
Again,
the $D$=3 data show a higher power level in this region,
indicative of more significant structure.
Note that there is a correlation in physical scales 
between the large power regions
in the father function representation and the fastest-changing distribution
widths in the mother function representation.
This is a manifestation of the relationship in Eq.\ (\ref{EQW})
between father- and mother-function expansions.

In the preceding,
we have presented analyses of various scenarios
that roughly emulate what could be expected
if the system undergoes a rapid non-equilibrium expansion,
as is the prerequisite for the occurrence of the \DCC\ phenomenon.
We have shown how suitable analyses may help to bring out
the associated anomalous pion multiplicity distribution
in quantities that are in principle observable.
However,
before closing we wish to address the issue of uniqueness:
would the appearance of an anomalous neutral fraction distribution $P(f)$
be a reliable \DCC\ signal?

In order to elucidate this central issue,
we have made similar analyses of events drawn
from equilibrium ensembles.
Figure \ref{fig:Pf-T} shows the neutral fraction distribution
extracted for the cubic sources in thermal equilibrium,
at temperatures of either $T$=100~MeV or $T$=200~MeV
(see fig.\ref{fig:Pf-cube}).
The results are practically indistinguishable
from the distributions obtained above with $D$=3 and $D$=2, respectively.
This remarkable result is due to the fact that the key quantity
determining the form of the extracted $P(f)$
is the pion correlation length.
In thermal equilibrium,
the correlation length grows steadily as the temperature is reduced and,
consequently, when analyzed at any given scale,
the associated neutral-fraction distribution becomes increasingly anomalous.
Thus, apparently to a quite good level of correspondence,
the sources from the non-equilibrium evolution
can be mocked up by equilibrium sources at a suitably adjusted temperature.
This approximate correspondence is also apparent in the scale dependence
of the  mother function distribution width in Fig.\ \ref{waveletwidth},
as well as in the wavelet power spectrum in Fig.\ \ref{waveletpower}.

It should be added that a more detailed analysis may help to resolve
this ambiguity,
because the correlation function resulting from non-equilibrium dynamics
differs qualitatively from the thermal form.
Whereas the latter exhibits a typical gaussian-like fall-off with distance
(see figs.\ 10-11 of Ref.\ \cite{JR:PRD}),
the former has an anomalous tail resulting from the amplified soft modes
(see Fig.\ 2 of Ref.\ \cite{JR:NPA}).
Consequently, in principle,
a sufficiently detailed analysis would reveal
whether the source is of thermal form.
In practice, though,
it might be difficult to acquire such detailed information.
It may then be necessary to carefully relate different types of observable,
including also electromagnetic probes:
photons \cite{bvhk} and dileptons \cite{zhxw96,kkrw97}.

\section{Concluding remarks}

The present study has been carried out for idealized scenarios
in order to better bring out the key features discussed.
Nevertheless,
it is possible to draw some lessons of direct observational relevance.

In particular,
our analyses suggest that it is important
to restrict the experimental acceptance
to small domains of suitable shape
so as to avoid the attenuation caused by the sampling of uncorrelated pions.
Thus, for example, in addition to binning the soft pion yields
according to rapidity
(and concentrating as far as possible on the soft end of the spectrum),
it would be preferable to also perform an azimuthal separation.
If one or more of these restrictions is relaxed,
the data collection will effectively sample different isospin domains
and thereby cause an attenuation of the anomalous behavior.

Our analyses also suggest that even under perfect experimental conditions,
one should not expect to observe the ideal $1/2\sqrt{f}$ form of the
neutral-fraction distribution $P(f)$,
since the isospin field is never fully aligned over a volume large enough
to contain several pions.
This inherent limitation underscores the need for the development
of suitable analysis methods to probe this type of physics.

Finally,
another important lesson that can be learned from the present study
is that the observation of a ``signal''
(in the form of an anomalous form of $P(f)$)
is by itself {\it not} sufficient evidence
that a catastrophic evolution has in fact occurred.
Indeed,
we showed that the distribution of the neutral pion fraction
resulting from a non-equilbrium process
can be well approximated by the distribution associated with full equilibrium,
at a suitable temperature.
Therefore, in order to disentangle this degeneracy,
it may be necessary to combine the measurements of pion distributions
with other observables,
such as electromagnetic probes.\\

This work was supported in part by the Director,
Office of Energy Research,
Office of High Energy and Nuclear Physics,
Nuclear Physics Division of the U.S. Department of Energy
under Contract No.\ DE-AC03-76SF00098.
We also wish to acknowledge the hospitality of the
National Institute for Nuclear Theory at the University of Washington
in Seattle where part of this work was carried out.


\newpage
\bfig[b]
\vspace{19cm}
\includegraphics{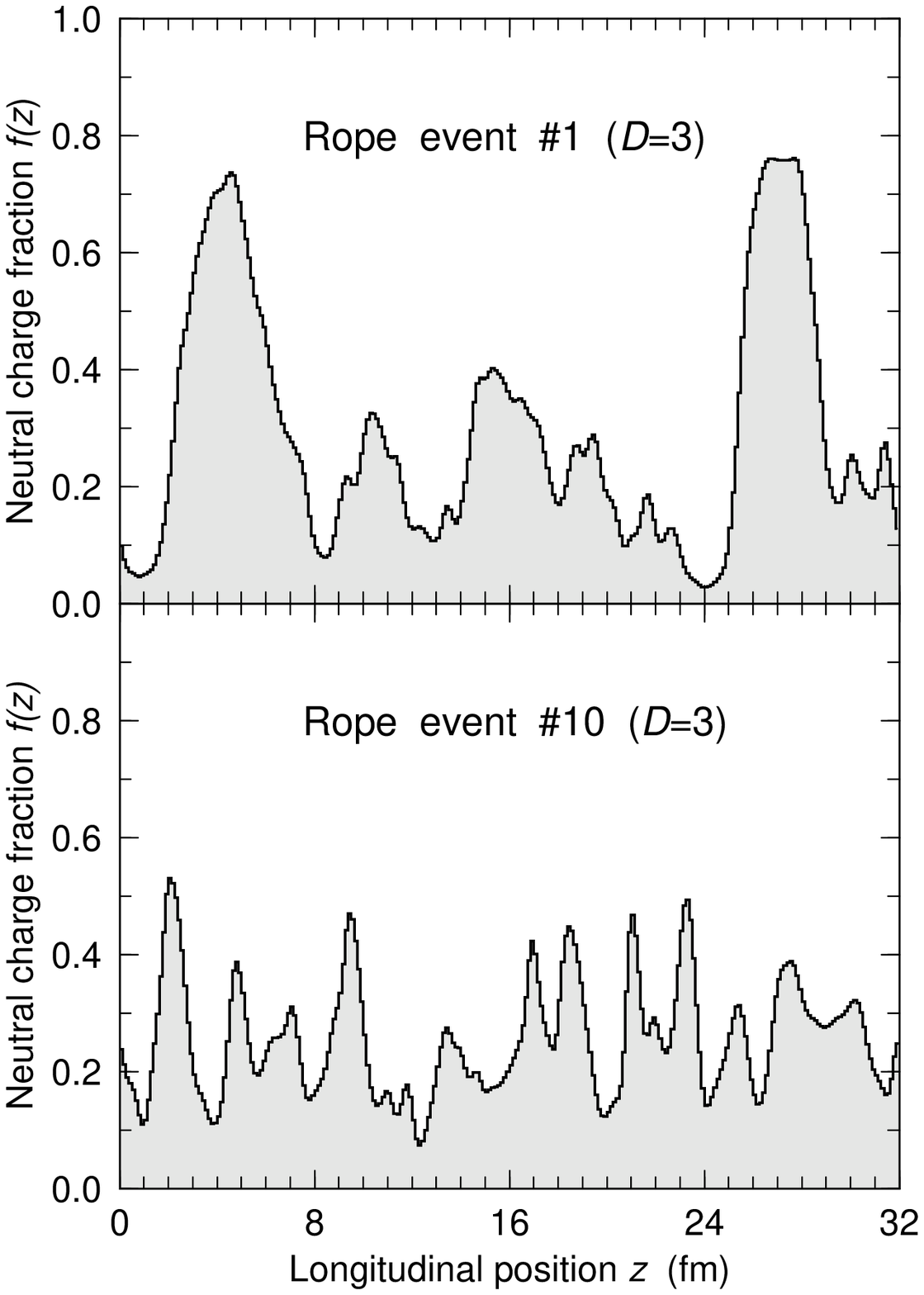}
\caption{Neutral pion fraction.}
\label{fig:f}
Histogram of the neutral pion fraction $f(z)=n_3(z)/n(z)$
for two rope configurations obtained in the rapid ($D$=3) cooling scenario.
The rope has a total length of 32 fm and a 4$\times$4 fm$^2$ cross section;
it has been cut into 256 slices, each having a thickness of $1\over8$ fm.
The abscissa is the corresponding longitudinal position $z$.
\efig

\newpage
\bfig[b]
\vspace{12cm}
\includegraphics{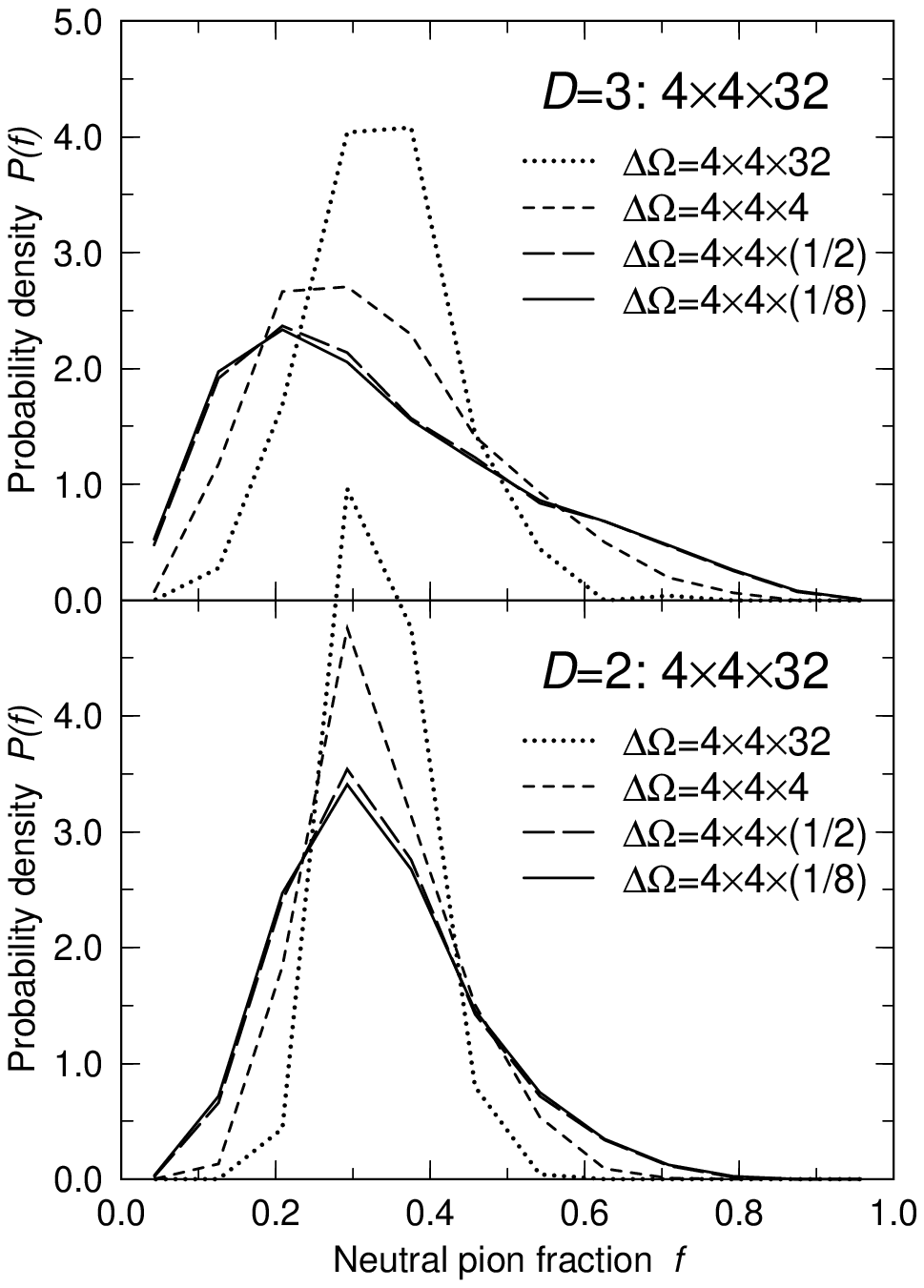}
\caption{Distribution of the neutral pion fraction for ropes.}
\label{fig:Pf-rope}
The distribution of the neutral pion fraction, $P(f)$,
as obtained by combining $2^{L-1}$ neighboring slices
in a sample of 100 rope configurations obtained with either $D$=2 or $D$=3.
The results are shown only for four different scales, $L=1,3,6,9$.
\efig

\newpage
\bfig[b]
\vspace{12cm}
\includegraphics{Fig3.ps}
\caption{Mother function analysis.}
The width of $P(f)$ as a function of the scale $L$
for the rope samples used in Fig.\ \ref{fig:Pf-rope} (solid curves),
configurations obtained by assigning random isospin directions 
in each basic slice (dots), 
and configurations sampled from thermal equilibrium ensembles
(dashed curves).
\label{waveletwidth}
\efig

\bfig[b]
\vspace{12cm}
\includegraphics{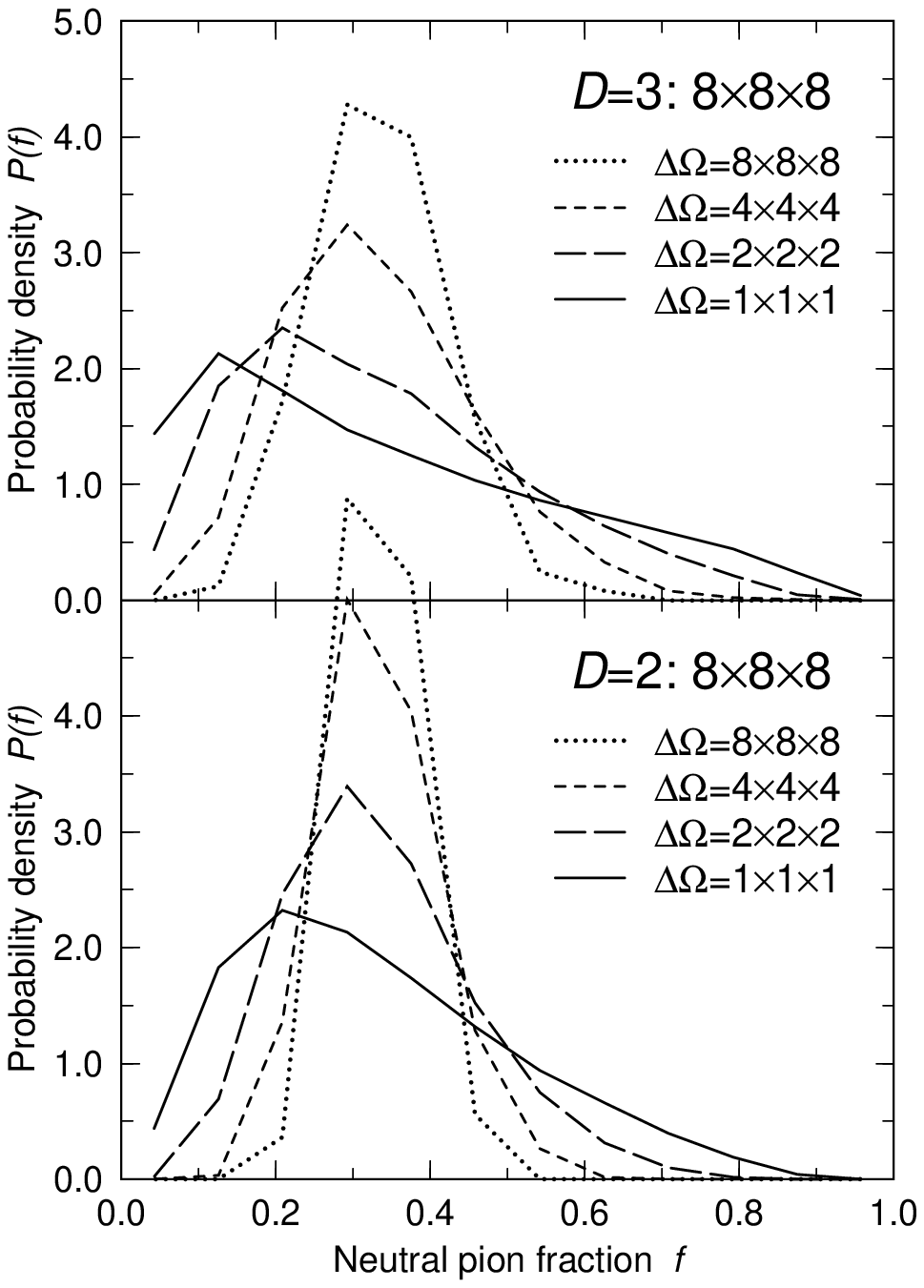}
\caption{Distribution of the neutral pion fraction for cubes.}
\label{fig:Pf-cube}
The distribution of the neutral pion fraction, $P(f)$,
as obtained by combining $2^L$ neighboring cubic cells
in a sample of 100 cube configurations obtained with either $D$=2 or $D$=3.
The results are shown only for four different scales, $L=0,3,6,9$,
so that the resulting sources all have a cubic shape.
\efig

\bfig[b]
\vspace{12cm}
\includegraphics{Fig5.ps}
\caption{Width of neutral fraction distribution.}
\label{fig:DfV}
The root-mean-square width $\Delta f$
of the neutral-fraction distribution $P(f)$
as a function of the scale $L$
(related to the source volume by $\Delta\Omega=2^L\ \fm^3$),
for both rope (solid) and cube (dashed) configurations
obtained with $D$=2 and $D$=3.
\efig

\newpage
\bfig[b]
\vspace{12cm}
\includegraphics{Fig6.ps}
\caption{Wavelet power spectrum.}
The power spectrum obtained with the {\sl DAUB4} wavelets
for samples of 100 configurations that have been obtained by
cooling with either $D$=3 or $D$=2 (solid),
assigning random field configurations at each grid point (dots),
or using a thermal source held at $T=100$, 140, 200 MeV (dashed curves).
\label{waveletpower}
\efig

\newpage
\bfig[b]
\vspace{12cm}
\includegraphics{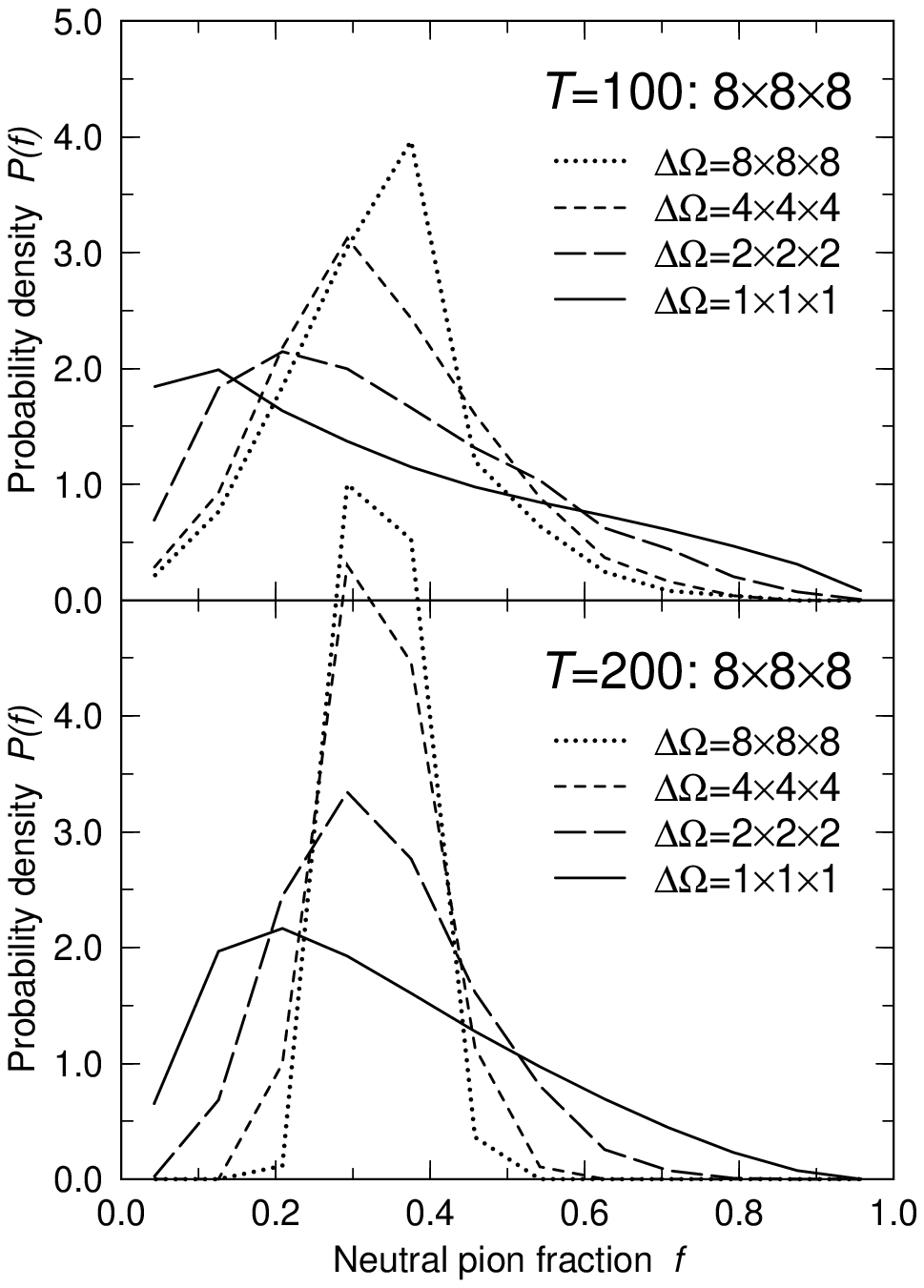}
\caption{Thermal fraction distribution.}
The distribution of the neutral pion fraction, $P(f)$,
obtained for rope and cube configurations
sampled from thermal equilibrium ensembles
at the specified temperatures $T$=100 MeV and $T$=200 MeV.
\label{fig:Pf-T}
\efig
			\end{document}